\documentclass[longauth]{aa} 
\usepackage{graphicx}
\usepackage{url}
\usepackage{color}
\usepackage{txfonts}
\usepackage{natbib}
\usepackage{ulem}
 \newcommand{\add}[1]{#1}
\newcommand{\rem}[1]{}

\newcommand{\remb}[1]{}

\newcommand{\addb}[1]{#1}
\begin{document} 

  \title{  The astrometric Gaia-FUN-SSO observation campaign of 99\;942 Apophis
\thanks{Table with positions and other data are available in electronic form at the CDS via anonymous ftp to cdsarc.u-strasbg.fr (130.79.128.5) or via  http://cdsweb.u-strasbg.fr/cgi-bin/qcat?J/A+A/??/A??}} 
   \author{Thuillot,~W.\inst{1}, Bancelin,~D.\inst{2, 1}, Ivantsov,~A.\inst{28,1}, Desmars,~J.\inst{4, 1}, Assafin,~M.\inst{3},  Eggl,~S.\inst{1,2},  Hestroffer,~D.\inst{1},  
Rocher,~P.\inst{1}, Carry,~B.\inst{1}, David,~P.\inst{1},
Abe,~L.\inst{10},
Andreev,~M.\inst{19},
Arlot,~J.-E.\inst{1},
Asami,~A.\inst{14},
Ayvasian,~V.\inst{25},
Baransky,~A.\inst{15},
Belcheva,~M.\inst{11},
Bendjoya,~Ph.\inst{10},
Bikmaev,~I.\inst{5,6}, 
Burkhonov,~O.A.\inst{26},
Camci,~U.\inst{7,30},
Carbognani,~A.\inst{17},
Colas,~F.\inst{1},
Devyatkin,~A.V.\inst{21},
Ehgamberdiev,~Sh.A.\inst{26},
Enikova,~P.\inst{11},
Eyer,~L.\inst{24},
Galeev,~A.\inst{5,6},
Gerlach,~E.\inst{20},
Godunova,~V.\inst{19},
Golubaev,~A.V.\inst{12},
Gorshanov,~D.L.\inst{21},
Gumerov,~R.\inst{5,6}, 
Hashimoto,~N.\inst{14},
Helvaci,~M.\inst{7},
Ibryamov,~S.\inst{11},
Inasaridze,~R.Ya\inst{25},
Khamitov,~I.\inst{7,5},
Kostov,~A.\inst{11},
Kozhukhov,~A.M.\inst{18},
Kozyryev,~Y.\inst{9},
Krugly,~Yu~N.\inst{12},
Kryuchkovskiy,~V.\inst{9},
Kulichenko,~N.\inst{9},
Maigurova,~N.\inst{9},
Manilla-Robles,~A.\inst{16, 29},
Martyusheva,~A.A.\inst{21},
Molotov,~I.E.\inst{27},
Nikolov,~G.\inst{11},
Nikolov,~P.\inst{11}, 
Nishiyama,~K.\inst{14},
Okumura,~S.\inst{14},
Palaversa,~L.\inst{24},
Parmonov,~O.\inst{26},
Peng,~Q.Y.\inst{8},
Petrova,~S.N.\inst{21},
Pinigin,~G.I.\inst{9},
Pomazan,~A.\inst{9},
Rivet,~J-P.\inst{10},
Sakamoto,~T.\inst{14},
Sakhibullin,~N.\inst{5,6},
Sergeev,~O.\inst{19},
Sergeyev,~A.V.\inst{12},
Shulga,~O.V.\inst{9},
Suarez,~O.\inst{10},
Sybiryakova,~Y.\inst{9},
Takahashi,~N.\inst{13},
Tarady,~V.\inst{19},
Todd,~M.\inst{22},
Urakawa,~S.\inst{14},
Uysal,~O.\inst{7,30},
Vaduvescu,~O.\inst{16,1},
Vovk,~V.\inst{9},
Zhang,~X.-L.\inst{23},
          }  

   \institute{
Institut de Mécanique Céleste et de Calcul des Éphémérides, Paris Observatory, UPMC, Lille 1 university, UMR 8028 du CNRS, PSL Research University, 77 avenue Denfert Rochereau 75014 Paris, France  (\email{thuillot@imcce.fr})
 \and
Institute for Astrophysics (IfA), University of Vienna, T\"urkenschanzstrasse 17, A-1180 Vienna, Austria
  \and
Observat\'orio do Valongo/UFRJ, Ladeira Pedro Antonio 43, CEP 20.080-090 Rio de Janeiro - RJ, Brazil
  \and
Observat\'{o}rio Nacional, Rua Jos\'e Cristino 77, S\~{a}o Cristov\~{a}o, Rio de Janeiro CEP 20.921-400, Brazil
  \and
Department of Astronomy and Geodesy, Kazan Federal University, Kremlevskaya Str., 18, Kazan, 420008, Russia
  \and
Academy of Sciences of Tatarstan, Bauman Str., 20, Kazan, 420111,Republic of Tatarstan, Russia
  \and
 TÜBİTAK National Observatory, Akdeniz University Campus, 07058, Antalya, Turkey
  \and
Department of Computer Science, Jinan University, Guangzhou 510632, P.R.China
   \and
Nikolaev Astronomical Observatory, 1 Observatorna, Mykolaiv, 54030, Ukraine
   \and
Université de Nice Sophia Antipolis, Observatoire de la Côte d'Azur, CNRS UMR 7293, Laboratoire Lagrange, Bd de l'Observatoire, B.P. 4229 06304 Nice Cedex 04 , France
   \and
Institute of Astronomy and NAO, Bulgarian Academy of Sciences, 72, Tsarigradsko Chaussee Blvd., 1784, Sofia, Bulgaria
   \and
Institute of Astronomy of Kharkiv National University, Sumska Street 35, Kharkiv 61022, Ukraine
   \and
Japan Spaceguard Association, 1-60-7 2F, Sasazuka, Shibuya-ku, 151-0073 Tokyo, Japan
   \and
Bisei Spaceguard Center, Japan Spaceguard Association, 1716-3 Okura,bisei, Ibara, 714-1411 Okayama, Japan 
   \and
Astronomical Observatory of Kyiv University, Observatorna street 3, Kyiv 04053, Ukraine
   \and
Isaac Newton Group of Telescopes, Apto. 321, 38700, Santa Cruz de la Palma, Canary Islands, Spain
   \and
Astronomical Observatory of the Autonomous Region of the Aosta Valley, Lignan 39, 11020 Nus (Aosta), Italy
   \and
Center of the Special Information Receiving and Processing and the Navigating Field Control, State Space Agency of Ukraine, Zalistsy, Dunayivtsy district, Khmelnytskyy region, 32444, Ukraine 
   \and
ICAMER Observatory, National Academy of Sciences of Ukraine, 27 Acad. Zabolotnoho Str., Kiev 03680, Ukraine
   \and
Technical University Dresden, Institute of Planetary Geodesy, Lohrmann Observatory, 01062 Dresden
   \and
Pulkovo Observatory, Pulkovskoe Ch. 65,196140 St.-Petersburg, Russia 
   \and 
Department of Imaging and Applied Physics, Bldg 301, Curtin University, Kent St, Bentley, WA 6102, Australia
   \and
Yunnan Observatories, Chinese Academy of Sciences (CAS), P. O. Box 110, Kunming 650011, China
   \and
Observatoire Astronomique de l'Université de Genève, 51 chemin des Maillettes, CH-1290 Sauverny, Switzerland
   \and
Kharadze Abastumani Astrophysical Observatory, Ilia State University, G.Tsereteli str. 3, Tbilisi 0162, Georgia
   \and
Ulugh Beg Astronomical Institute, Astronomicheskaya Street 33, Tashkent 100052, Uzbekistan
   \and
Keldysh Institute of Applied Mathematics, RAS, Miusskaya sq. 4, Moscow 125047, Russia
   \and
Faculty of Aerospace Engineering, Technion–Israel Institute of Technology, Technion City, 3200003 Haifa, Israel
   \and
Universidad de La Laguna (ULL), E-38205 La Laguna, Tenerife, Spain
   \and
Department of Physics, Science Faculty, Akdeniz University, 07058 Antalya, Turkey
\\
       }

   \date{Received December XX, 2014; accepted XXXXXX XX, XXXX}

  \abstract
   {}
  {Astrometric observations performed by the Gaia Follow-Up Network for Solar System Objects (Gaia-FUN-SSO) play a key role in ensuring that moving objects first detected by ESA's Gaia mission 
  remain recoverable after their discovery.  An observation campaign on the
  potentially hazardous asteroid (99~942) Apophis was conducted during the asteroid's latest period of visibility, from 12/21/2012 to 5/2/2013, to test the coordination and
evaluate the overall performance of the Gaia-FUN-SSO .
 }
{The 2732 high quality astrometric observations acquired during the Gaia-FUN-SSO campaign were reduced with the Platform for Reduction of Astronomical Images Automatically (PRAIA), 
using the USNO CCD Astrograph Catalogue 4 (UCAC4) as a reference. 
The astrometric reduction process and the precision of the newly obtained measurements are discussed.
We compare the residuals of astrometric observations that we obtained using this reduction process  to data sets that were individually reduced by observers and accepted by the Minor Planet Center.
}
   {We obtained 2103 previously unpublished astrometric positions  and   provide these to the scientific community.
   Using these data we show that our reduction of this astrometric campaign with a reliable stellar catalog substantially improves the quality of the astrometric results.
We present evidence that the new data will help to reduce the orbit uncertainty of Apophis during its close approach in 2029.
 We show that uncertainties due to geolocations of observing stations, as well as rounding of astrometric data can introduce an unnecessary 
degradation in the quality of the resulting astrometric positions. Finally, we discuss the impact of  our campaign reduction on the recovery process of newly discovered asteroids.
}  
   {}


\titlerunning{The astrometric Gaia-FUN-SSO observation campaign of (99\;942) Apophis}
\authorrunning{Thuillot W. et al.}

\maketitle
\section{Introduction}
  Recent decades have seen a growing scientific and social interest
  toward near-Earth objects (NEOs) and, especially, near-Earth asteroids (NEAs). 
  Like their former associates in the main belt, NEAs are remnants
  from the early solar system having experienced relatively little physical
  evolution, apart from minor collisions and space weathering. This makes them a
  good tracer of the early stages of our solar system's formation. 
  Moreover, being  close to Earth, these objects are good candidates
  for space missions aimed at boosting our understanding of these
  objects' composition. Also, NEAs  pose a potential risk when impacting the
  planet. Unfortunately, we are still far from having catalogued the 
  entire NEO population. In fact,
  current estimates only give a satisfying level of completeness  for
  objects larger than 1\,km ($>90$\%). This percentage rapidly decreases for smaller
  objects, however. Recent results from the NEOWISE mission \citep{Mainzer12}
  have helped to revise the number of objects larger than 100\,m leading to a 
  current level of completeness of roughly $30$\%, while
  \citet{Brown13} claim that the number of smaller objects is one
  order of magnitude larger for diameters between 10-50\,m reducing the  
  corresponding level completeness to merely $3$\%. Detecting objects
  throughout the whole NEOs population is, thus, still important.
  Attaining information about their composition and constraining orbit uncertainties of 
  newly discovered as well as known objects is just as vital.  
  Indeed, openly discussing the potential threat that arises from a collision of an asteroid or comet with the Earth
  has raised awareness in the international community that  
  it is desirable to have the capabilities of predicting future impacts. In Europe, the \addb{Space Situation Awareness (SSA)} program of 
  \addb{European Space Agency (ESA)} is developing activities together with \addb{European Union (EU)} and the NEOshield project
  \citep{Harris12_epsc}. Observational efforts are also performed by
  the Euronear consortium \citep{Vaduvescu13}.  
  Other programs are developed at a global level, such as
  the United Nations Committee on the Peaceful Uses of Outer Space
  action team 14  (UN-COPUOS AT-14), the International Asteroid Warning Network (IAWN)\footnote{http://www.minorplanetcenter.net/IAWN/},
   European Space Agencies' Space Mission Planning Advisory Group (SMPAG)\footnote{http://www.cosmos.esa.int/web/smpag}, and the International Astronomical Union (IAU) Working Group on 
  \addb{NEOs}. 
  While the risk of collision with large bodies (diameters >1\;km) 
  within the next century is currently considered to be small, the impact of  smaller
  objects should be a major current concern.  
  The recent Chelyabinsk event in February 2013 has shown that objects
  of several tens of meters present a moderate yet real risk to cities,
  since they are difficult to track with current means.
  This results in short to nonexistent warning times.
  To detect and catalog a larger fraction of NEOs, several surveys have been set up,
  such as Lincoln Near-Earth Asteroid Research (LINEAR), Catalina, the most recent one being the Panoramic Survey Telescope \& Rapid Response System, \addb{PanSTARRS}  \citep{Jedicke07_dps}. 
  Similarly, the Gaia mission, successfully launched  at the end of
  2013, will continuously scan the sky over five years, observing stars,
  QSOs, and galaxies, as well as several 100,000s small bodies of the
  solar system \citep{mignard07,hestro10_lnp}. Since Gaia is operating
  in space down to solar elongations as low as 45~degrees,  
  it has the possibility to discover asteroids at low elongation and
   is better at probing the \addb{inner-Earth dynamical region, where Atira
    asteroids orbit}. 
  A network of ground-based telescopes has been set up (the Gaia-FUN-SSO, standing
  for Gaia Follow-Up Network for Solar System Objects) to
  avoid the rapid loss of a newly discovered fast-moving object because of
Gaia's lack of follow-up capabilities. The aim of the
  network is to recover, on alert, these asteroids and provide required
  astrometry for preliminary orbit estimation. 
Up to the date of this article no alert is yet triggered from Gaia, but the network is ready to operate. Thereby we could organize several campaigns of observation of NEOs both for contributing to NEO science and for testing the network.

Among all catalogued NEOs, some have been classified as potentially hazardous asteroids (PHA) because of their small distance to Earth orbit (MOID) and estimated size, and these  require particular attention. 
The asteroid (99~942) Apophis is one of those classified as potentially hazardous.
Considering the interest in obtaining a dense observational coverage of the PHA Apophis, to better constrain its dynamics, we organized an observation campaign in the
framework of the Gaia-FUN-SSO during the previous period of visibility of the asteroid. 

\add{Given the large number of participants, this campaign offered a good opportunity to test the efficiency of the astrometric data gathering process in the Follow-Up Network. 
Here we review the operations of the Gaia-FUN-SSO during the Apophis campaign in 2012/2013 (sections \ref{sec:apophis}  and \ref{sec:campaign}) 
and present previously unpublished astrometric observations of (99~942) Apophis (section \ref{sec:data}). 
 Furthermore, we had  the opportunity to perform a homogeneous treatment of the collected CCD images with the Platform for Reduction of Astronomical Images Automatically (PRAIA) software described by \citet{assafin2011}, using the USNO CCD Astrograph Catalogue 4 (UCAC4). In section (\ref{sec:homogeneous}) we discuss whether astrometric positions can be improved by this procedure. 

The influence of astrometric data improvement on the predictability of newly discovered objects is discussed in section \ref{sec:alerts}.
In section \ref{sec:bplane} we study the impact of the new astrometric data on the position uncertainty of Apophis during its close approach
to the Earth in 2029.
Finally, sections \ref{sec:geoposition} and \ref{sec:rounding} contain a detailed discussion of adverse effects due to the imprecise 
geolocations of observatories as well as the MPC format based rounding of astrometric data. 
Our findings are summarized in section \ref{sec:conclusions}.
}

\section{The PHA (99~942) Apophis}
\label{sec:apophis}
Ever since its discovery in 2004 by R.A.~Tucker, D.J.~Tholen, and F.~Bernardi, the asteroid (99~942) Apophis (2004 MN$_4$) has been a cause for concern.
Having one of the highest impact probabilities in the currently known \addb{near-Earth asteroid} population
\footnote{NASA JPL SENTRY: http://neo.jpl.nasa.gov/risks/, 
NEODyS: http://newton.dm.unipi.it/neodys/index.php?pc=4.0 }, 
it was named after the Egyptian god of chaos. If the $375^{+14}_{-10}$\,m diameter NEA \citep{muller-2014} were to reach the
surface of the Earth, it could deposit roughly between 750 and 1430\,Mt of TNT during an impact\footnote{Considering the range of relative
velocities during close encounters of Apophis with the Earth from 1907 to 2029, we find the likely top of the atmosphere
impact energies of $1020_{-270}^{+410}$\,Mt of TNT,
assuming that the mass range provided by \citet{muller-2014} is accurate.}
causing widespread devastation and a substantial loss of human life.

Fortunately, a collision between Apophis and the Earth can be ruled out for the foreseeable future \citep{bancelin-2012,farnocchia-2013}.
However, Apophis' trajectory remains difficult to predict because of  a deep close encounter with the Earth in 2029.
In fact, on April 13, 2029 Apophis will pass by the Earth at such a small
distance \addb{(only 5-6 Earth radii)}
that it will be observable to the naked eye from several countries. 

As the asteroid's trajectory can be significantly changed
during this event, even small uncertainties attributed to the asteroid's state vector tend to have a huge impact
on collision probabilities after 2029 \citep{chesley-2006}. \citet{giorgini-2008} have shown, for instance, that even tiny nongravitational
effects, such as the Yarkovsky drift, can significantly influence impact predictions.
Even though recent studies on the current and future spin states of Apophis indicate that the acceleration due to the Yarkovsky effect is
most likely smaller than originally presumed \citep{pravec-2014,lhotka-2013,scheeres-2005}, its influence remains non-negligible.
\citet{chesley-2006} and \citet{giorgini-2008} pointed out the importance of past and future radar observations
for reducing the uncertainties in Apophis' orbit and drift parameters. \citet{bancelin-2012} and \citet{farnocchia-2013}
also studied the impact of astrometric measurements in this respect. Their results indicate that high quality astrometric data
can make a substantial contribution to improving NEA impact predictions by reducing orbit uncertainties even when radar observations are available.

\add{In the following, we  discuss the treatment of astrometric data acquired by the Gaia-FUN-SSO network during the close approaches of
Apophis in 2012/2013\addb{, which is a typical example of a set of
  astrometric observations obtained by numerous observers.}
We  show that our analysis can decrease systematic errors and boost the quality of astrometric positions.}


\section{The Gaia-FUN-SSO observing campaign}
\label{sec:campaign}

  The Gaia mission, launched on 19 December 2013, is currently mapping
  the sky during five years and is performing astrometric measurements
  at an unprecedented level of precision (down to 9~$\mu$arcsec for
  stellar objects\footnote{Website
    http://www.cosmos.esa.int/web/gaia/science-performance}, down to 1~mas for solar system objects). The probe is on a Lissajous orbit
  around the L2 Lagrange point of the Sun-Earth system. It is spinning
  around its axis, which precesses around the direction to the Sun with
  an angle of 45 degrees. Gaia will continuously scan the sky according
  to a specific scanning law \citep{mignard-2007}.  
  In the case that new solar system objects are detected, an alert mode has been set up  to identify these
  objects and trigger complementary observations from the ground, since the probe cannot keep monitoring its
  discoveries. 

  To deal with alerts, a ground-based follow-up network has been developed \citep{thuillot2011} on the basis 
  of registration of observatories\footnote{Website
    http://gaiafunsso.imcce.fr} volunteering to participate in the follow-up of solar system objects. 
  To date, Gaia-FUN-SSO encompasses 57 observing sites operating 79 telescopes. 
  A central node located at IMCCE/Paris Observatory manages Gaia-FUN-SSO. 
  When discovery alerts are triggered by one of the Gaia data centers located at the French space agency CNES (Toulouse, France), they are ingested into a pipeline for publication and dissemination to the network. 
  Observers are in charge of the retrieval of the SSO they have been assigned. They also carry out the astrometric measurements and send them to the Minor Planet Center. 
  Thanks to periodic updates of the auxiliary database used by Gaia for identifying new objects, the improved orbital
  elements of the discoveries are systematically taken into account. 
    
  To ensure compliance with Gaia-FUN-SSOs astrometric standards, specific training campaigns have been organized over the past three years. In particular, 
  the close approach of (99~942) Apophis was a fine opportunity for both training observers and collecting useful data. 
  Table~\ref{T:obs} gives some characteristics of the
  observing sites that participated in this campaign. This campaign
  of observations allowed collection of extensive observations from
  which 2732 valuable astrometric measurements were extracted.  The
  corresponding observation arc ranges from
  \remb{21/12/2012 to 2/5/2013}
  \addb{12/21/2012 to 5/2/2013}
  (see Fig.~\ref{F:Dcosd-Dd-time}). 
  
  Some of the observations performed have been reduced by the observers themselves,  using their preferred tools and astrometric catalogs. Those results were submitted to the MPC. 
  However, we decided to conduct a complementary, homogeneous reduction, 
  with all CCD images recorded during this campaign using the PRAIA reduction pipeline \citep{assafin2011} and the UCAC4 astrometric catalog \citep{zacharias2013}.  

   \begin{table*}[h]
      \caption[] {Observatories of the Gaia-FUN-SSO involved in the Apophis campaign.}
          \label{T:obs}
          \begin{tiny}
      $$ 
            \begin{tabular}{ccrrrcccc}
             \hline
             \noalign{\smallskip}
            MPC &  Telescope / MPC Observatory Name\tablefootmark{1} &  Long.\tablefootmark{2} (E) & Lat.\tablefootmark{2} (N) & Height & Telescope & FOV & Pixel   \\
            code &   &  deg. & deg. & m &  diam., m & arcmin & arcsec. \\

             \noalign{\smallskip}
             \hline
              \noalign{\smallskip}
           010 & C2PU / Caussols                                                  & 6.92272 & 43.75374 & 1263 & 1.00 & $12\times12$ & 0.17 \\
           071 & Schmidt / NAO Rozhen, Smolyan                         & 24.73878 & 41.69725 & 1749 & 0.50 & $74\times74$ & 1.08\tablefootmark{3} \\
           089 & Mobitel / Nikolaev                                                  & 31.97358  & 46.97114 & 49 & 0.50  & $43\times21$ & 0.84   \\
           119 & Meniscus AS-32 / Abastuman                               & 42.81940  & 41.75404  & 1584 & 0.70  & $44\times30$  & 0.87 \\
          188 & AZT-22 / Majdanak                                                 & 66.89573  & 38.67345 & 2588 & 1.50 & $11\times11$  & 0.21 \\
           300 & Bisei Spaceguard Center-BATTeRS                      & 133.54527 & 34.67193 & 418 & 1.00  & $73\times36$  & 2.11 \\
           511 & T120 / Haute Provence                                          & 5.71513     & 43.93176 & 638 & 1.20  & $12\times12$  & 0.68 \\
           585 & Kyiv comet station                                                 & 30.52464   & 50.29786 & 146 & 0.70  & $16\times17$  & 0.95 \\
           586 & T1m / Pic du Midi                                                  & 0.14279    & 42.93644 & 2877 & 1.05  & $8\times8$  & 0.44 \\
           950 & William Herschel Telescope / La Palma                 & $-17.87757$ & 28.76212 & 2337 &  4.20  & $9\times10$ & 0.25 \\
           A84 & T100 / TUBITAK National Observatory                   & 30.33575  & 36.82156 & 2473 & 1.00  & $21\times21$ & 0.31 \\
           A84 & RTT150 / TUBITAK National Observatory              & 30.33553  & 36.82563 & 2462 & 1.50 &  $13\times13$ & 0.39 \\
           B04 & OAVdA, Saint Barthelemy                                      & 7.47853     & 45.78975 & 1668 & 0.81 & $16\times16$ & 0.96 \\
           B17 & AZT-8 Evpatoria                                                     & 33.16286   & 45.21949 & 12 & 0.70  & $45\times45$  & 1.76 \\
           B18 & Zeiss-600 / Terskol                                                 & 42.50047   & 43.27499 & 3143 & 0.60 & $11\times11$  & 1.24 \\
           C01 & Lohrmann-Observatorium, Triebenberg                 & 13.92293   & 51.02727 & 380 & 0.60 & $51\times51$ & 0.75 \\
           C20 & Kislovodsk Mtn. Astronomical Stn., Pulkovo Obs.  & 42.66297   & 43.74156 & 2063 & 0.50 & $20\times20$   & 1.19 \\
           D20 & Zadko Observatory, Wallingup Plain                       & 115.71317 & $-31.35594$ & 51 & 1.00 & $24\times24$  & 1.38 \\
           O44 & Lijiang Station, Yunnan Observatories                   & 100.02985 & 26.69504 & 3230 & 2.40 & $10\times10$   & 0.28 \\
           Z20 & La Palma - Mercator                                                & $-17.87847$ & 28.76237 & 2333 & 1.20 & $10\times14$ & 0.55 \\
            \noalign{\smallskip}
            \hline
          \end{tabular}
      $$
      \end{tiny}
      \tablefoot{
      \tablefoottext{1}{There can be several telescopes within one observatory used for observations of asteroids. Hence, we specify the particular telescope used during the Apophis campaign.}
      \tablefoottext{2}{Longitude and latitude coordinates are given with respect to the World Geodetic System 1984 (WGS84). They can be verified using the Google Earth software, https://www.google.com/earth/. The last given digits are uncertain.}
      \tablefoottext{3}{Different CCD binning modes were used for some observations obtrained with this telescope. When binning was applied the pixel size increased to $2.16\arcsec$.}}
    \end{table*}


\section{Astrometric Reduction}
\label{sec:data}
Prior to any astrometric reductions, the frames were \rem{photometrically
calibrated}{corrected} with auxiliary bias and flat-field frames by means of
standard procedures using IRAF\footnote{Website:
  http://iraf.noao.edu/}, whenever such data were available.
The astrometric reductions were then performed using PRAIA. Although PRAIA is capable of providing photometric data as well, 
we  focus on astrometric image reduction. 
 Exposure times varied mostly from 30 s to 120 s and the seeing from 1.2 to 4 arcsec. Consequently the trailing was negligible so that 
the (x, y) \addb{positions on the CCDs could be
  measured} \rem{measurements were performed}with two-dimensional
circular symmetric Gaussian fits. The (x,y) errors ranged from 10 mas up to 30 mas. We chose the UCAC4
\citep{zacharias2013} as the practical representative of the
International Celestial Reference System (ICRS). Standard
bivariate polynomials were used to model the (x, y) measurements to the (X, Y)
tangent plane coordinates. Depending on the size and distortions in
the FOV of each observatory/telescope image set, the six constants
model up to a complete polynomial of the third degree. One by
one, outlier reference stars were eliminated in an iterative reduction
procedure until all absolute values of the stellar position residuals were below 120 mas. 
The latter number results from pessimistic estimates of the UCAC4 catalog error \citep{Zacharias04,Zacharias10, Zacharias12}.
No weights were used for the reference stars. 

\begin{figure}[h!]
\begin{tabular}{c}
\includegraphics[angle=-90, width = \columnwidth ]{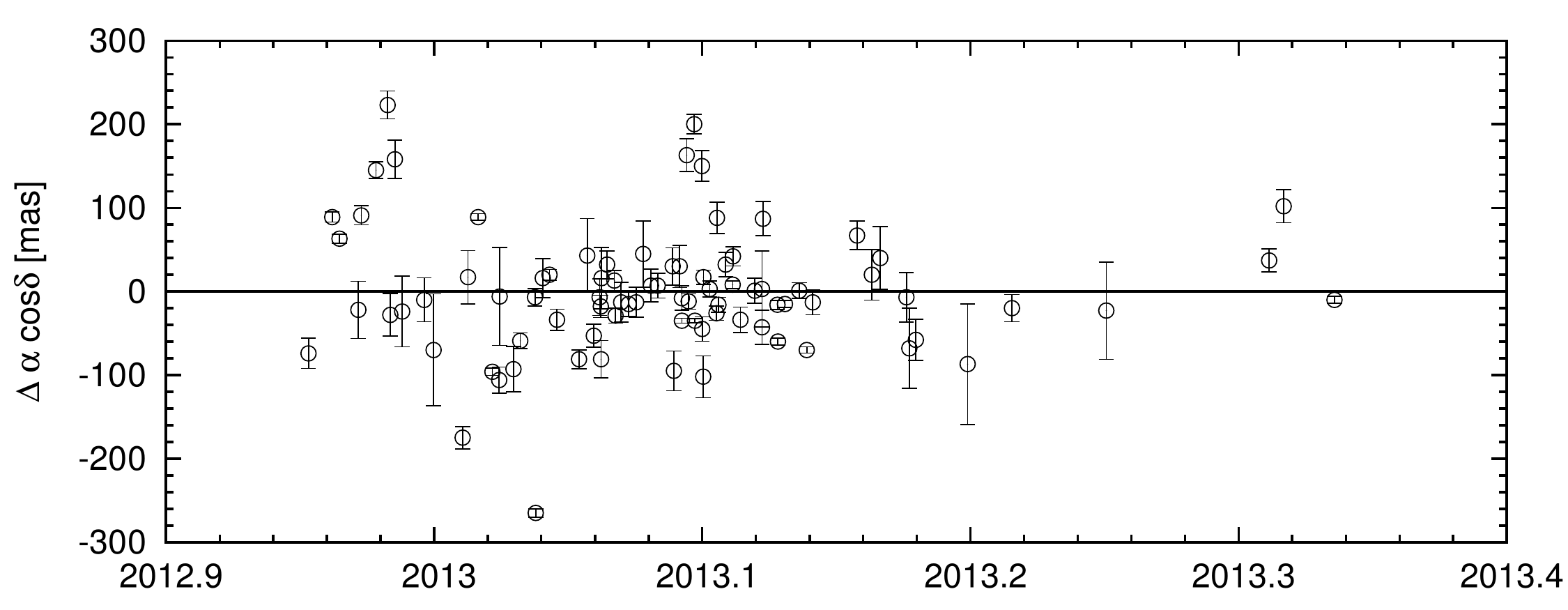}\\
\includegraphics[angle=-90, width = \columnwidth]{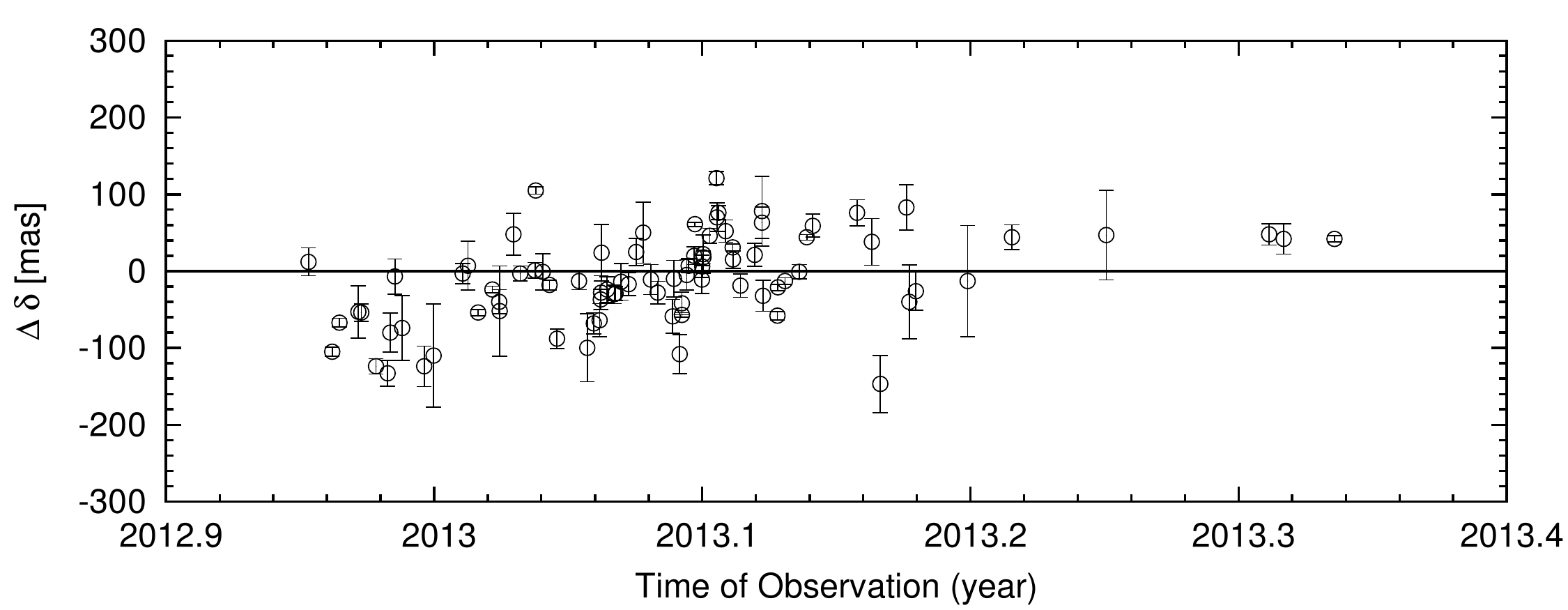}
\end{tabular}
\caption{Nightly averages and 1 sigma uncertainties of (O-C)s for the asteroid (99~942) Apophis versus time;
O: PRAIA astrometric right ascension and declination, C: JPL193/DE431 derived astrometric positions.}
\label{F:Dcosd-Dd-time}
\end{figure}

\begin{figure}[h!]
 \centerline{
\includegraphics[ angle=-90,width = \columnwidth]{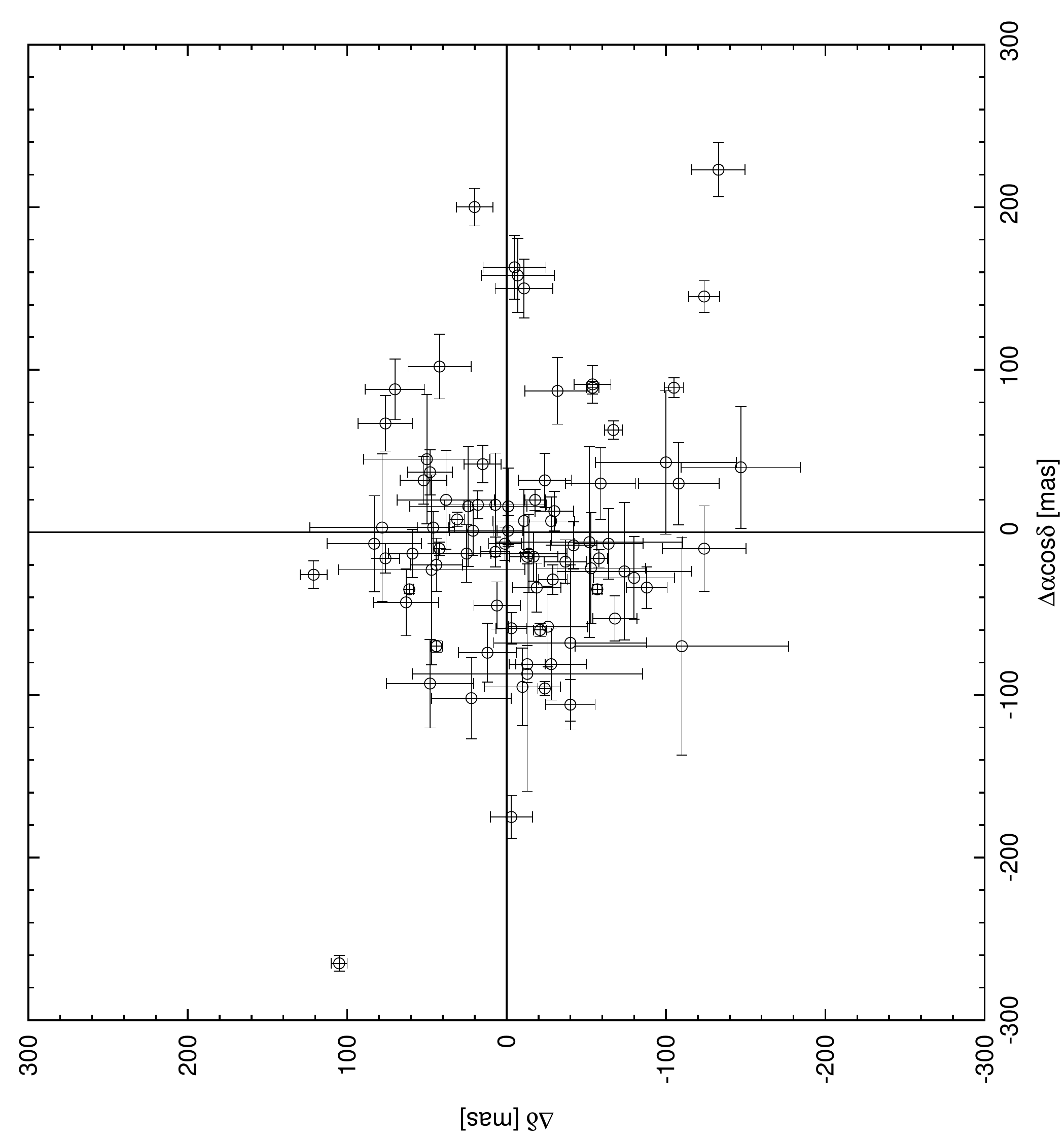}
}
\caption{Scatter plot of nightly averages and 1 sigma uncertainty bars of (O-C)s for the asteroid (99~942) Apophis; 
O: PRAIA astrometric right ascension and declination, C: JPL193/DE431 derived astrometric positions.}
\label{F:Dcosd-Dd}
\end{figure}

Table~\ref{T:reduction} lists the astrometric information from the
reductions for each telescope set. The
mean errors in the frame stars' right ascension and declination ($\alpha$, RA) and ($\delta$, DEC) with respect to catalog positions are given in the last two columns. 
The average number of reference stars per frame, the number of nights, and
 total number of usable positions are also provided. Table~\ref{T:reduction} furthermore lists the ($\alpha$,
$\delta$) standard deviations from the nightly average offsets with
respect to JPL193/DE431 ephemerides, mean values $\langle O-C \rangle_n$, after the removal of statistical outliers.  
Comparing the scattering of the residuals $\langle O-C \rangle_n$
in Figs. \ref{F:Dcosd-Dd-time}  and \ref{F:Dcosd-Dd} 
to the recently improved post-fit residuals for 
Apophis \citep[][Fig. 20]{farnocchia-2015},
one can see that the  presented positions are of a quality similar 
to the measurements selected by JPL to produce Apophis current orbit (JPL193/DE431). 

  The final reduction yields a consistent set of 2732 astrometric
  measurements of Apophis, which have been formatted according to the
  MPC\footnote{see http://www.minorplanetcenter.net/iau/info/OpticalObs.html}  with extra precision in the epoch of observations ($10^{-6}$ days), 
  right ascension ($0.001$s), and
  declination ($0\arcsec.01$). Of those observations, 2103 have not been published previously.  
  Table~\ref{T:newastrometry}
  shows a sample of the data accompanying this article.


\begin{table}[]
\caption{\label{T:reduction} Astrometric information on the PRAIA reduced ($\alpha$, $\delta$) sets for each observatory.}
\begin{centering}
\begin{tiny}
\begin{tabular}{lccccccc}
\hline\hline
IAU&\multicolumn{2}{c}{S. D. to JPL}& No. & No. & UCAC4 & \multicolumn{2}{c}{RMS to Cat.} \tabularnewline
code&$\sigma_{\alpha}\cos\delta$ & $\sigma_{\delta}$&nights&pos.&stars&$\sigma_{\alpha}cos\delta$&$\sigma_{\delta}$\tabularnewline
&  mas & mas & & & & mas & mas\tabularnewline
\hline
010 & 48 & 49 & 1 & 137 & 94 & 61 & 63 \tabularnewline
071 & 96 & 80 & 6 & 114 & 1336 & 57 & 56 \tabularnewline
089 & 102 & 137 & 4 & 74 & 540 & 56 & 62 \tabularnewline
119 & 81 & 57 & 2 & 7 & 621 & 59 & 58 \tabularnewline
188 & 83 & 50 & 1 & 22 & 20 & 52 & 52 \tabularnewline
300 & 63 & 80 & 4 & 13 & 528 & 60 & 60 \tabularnewline
511 & 40 & 46 & 1 & 7 & 83 & 56 & 53 \tabularnewline
585 & 74 & 71 & 3 & 15 & 180 & 57 & 56 \tabularnewline
586 & 100 & 75 & 6 & 960 & 36 & 63 & 66 \tabularnewline
950 & 40 & 23 & 1 & 5 & 24 & 72 & 50 \tabularnewline
A84\tablefootmark{1} & 55 & 27 & 1 & 124 & 224 & 52 & 50 \tabularnewline
A84\tablefootmark{2} & 64 & 58 & 5 & 30 & 145 & 57 & 55 \tabularnewline
B04 & 172 & 283 & 2 & 16 & 110 & 60 & 62 \tabularnewline
B17 & 51 & 90 & 6 & 22 & 738 & 63 & 64 \tabularnewline
B18 & 71 & 82 & 4 & 126 & 71 & 60 & 58 \tabularnewline
C01 & 86 & 144 & 2 & 7 & 1822 & 58 & 62 \tabularnewline
C20 & 62 & 68 & 18 & 664 & 210 & 61 & 62 \tabularnewline
D20 & 85 & 68 & 22 & 147 & 247 & 62 & 56 \tabularnewline
O44 & 33 & 49 & 4 & 70 & 102 & 58 & 60 \tabularnewline
Z20 & 56 & 57 & 4 & 160 & 16 & 62 & 62 \tabularnewline
\hline
\end{tabular}
\end{tiny}
\par\end{centering} 
\tablefoot{ S. D. to JPL denotes ($\alpha$, $\delta$) standard deviations about the nightly average with respect to the JPL reference ephemeris, 
after the elimination of outliers. The values in the RMS to catalog columns are root mean square residuals from the field stars' ($\alpha$, $\delta$) positions with respect to UCAC4.
Detailed telescope data for each observatory are given in Table \ref{T:obs}.
\tablefoottext{1}{Observations at the T100 telescope.}
\tablefoottext{2}{Observations at the RTT150 telescope.}}
\end{table}


 \begin{table*}
\caption {New astrometric measurements of Apophis, according to the MPC format, with three digits for RA seconds and two digits for DEC arcseconds. This is a sample; the full table, including the X Y positions of the asteroid and the reference stars, is accessible at CDS.}
\label{T:newastrometry}
\begin{tiny}
      $$ 
          \begin{tabular}{p{0.03\linewidth}ccccc}
             \hline
             \noalign{\smallskip}
            Asteroid &        &   Date          &      RA               &      DEC                   & Obs code   \\
             number &        &   y m d         &   h       m       s   &   deg.  '  "               &      \\    
             \noalign{\smallskip}
             \hline
              \noalign{\smallskip}
99942&         &C2013 02 15.950590&06 48 26.290&+01 12 28.68&            010\\
99942&         &C2013 02 15.951076&06 48 26.248&+01 12 29.69&            010\\
99942&         &C2013 02 15.951563&06 48 26.211&+01 12 30.66&            010\\
99942&         &C2013 02 15.952535&06 48 26.142&+01 12 32.77&            010\\
99942&         &C2013 02 15.953021&06 48 26.107&+01 12 33.85&            010\\
99942&         &C2013 02 15.953519&06 48 26.070&+01 12 34.80&            010\\
99942&         &C2013 02 15.954005&06 48 26.035&+01 12 35.86&            010\\
99942&         &C2013 02 15.954491&06 48 26.006&+01 12 36.87&            010\\
99942&         &C2013 02 15.954977&06 48 25.966&+01 12 37.88&            010\\
99942&         &C2013 02 15.955463&06 48 25.933&+01 12 38.95&            010\\
99942&         &C2013 02 15.955949&06 48 25.896&+01 12 39.96&            010\\
99942&         &C2013 02 15.956447&06 48 25.858&+01 12 41.00&            010\\
99942&         &C2013 02 15.956933&06 48 25.820&+01 12 42.00&            010\\
99942&         &C2013 02 15.957419&06 48 25.784&+01 12 43.03&            010\\
.../... \\
            \noalign{\smallskip}
            \hline
          \end{tabular}
      $$ 
\end{tiny}
    \end{table*}    
    
\section{\addb{Homogeneous vs heterogeneous reduction}\remb{Data analysis}}
\label{sec:homogeneous}
One of the main questions we would like to answer in this article is the following: 
Can a homogeneous reduction pipeline improve the quality of astrometric data sets? 
We tackle this question using comparative statistics.
Among the 2732 astrometric observations that result frorm the Gaia-FUN-SSO campaign, 629 had already been published at the MPC. 
Three of those 629 were later  identified as likely outliers and had to be rejected.
In other words, 626 high quality observations are available, which  have been reduced in parallel 
using PRAIA and UCAC4, on the one hand, and different reduction software and catalogs, on the other hand. 
We, thus, define the following observation sets for later reference:
\begin{itemize}
 \item D$_{\scriptscriptstyle MPC}$ refers to the 626 duplicated Gaia-FUN-SSO astrometric measurements already sent to the MPC by
the observers. The corresponding observations were reduced with various astrometric software packages and catalogs.
\item D$_{\scriptscriptstyle PRAIA}$ refers to the same 626 Gaia-FUN-SSO observations, but re-reduced with PRAIA using the UCAC4 astrometric catalog.
\end{itemize}
While all astrometric reduction with PRAIA was done using UCAC4, 
the reductions that had been performed by the observers themselves were based on various catalogs, such as \mbox{NOMAD}, USNO-B1.0, PPMXL, UCAC2, UCAC3, and UCAC4.
Different reduction  software suites were used as well. 
Unfortunately, there were also instances where the name of the catalog used for reduction was not reported to the MPC. 
As the  quality of catalogs is not necessarily the same \citep{chesley-2010,farnocchia-2015}, 
we  investigate the effect that open choice of the reduction pipeline has on the overall data quality.
To avoid being dependent on a specific ephemeris service to conduct this comparison, 
we  only used statistics on the astrometric position measurements of Apophis.
There  several related questions emerge about this.
\begin{itemize}
\item How does the choice of the reduction software influence astrometric results when the same catalogs underlie the analysis? 
\item Are the limitations of different catalogs traceable in the astrometric reduction of Apophis observations?
\item How large is the combined uncertainty of software and catalogs on the astrometric positions of asteroids?
\end{itemize}
We form three subsets of $D_{MPC}$, i.e., the 626 observations reduced by the observers themselves.
\begin{enumerate}
\item \textbf{\textit{R}}: 152 positions reduced using the UCAC4 catalog
\item \textbf{\textit{U}}: 457 positions reduced using any UCAC catalog, i.e., UCAC2, UCAC3, and UCAC4.
\item \textbf{\textit{O}}: 169 positions reduced using non-UCAC or nondescript catalogs. 
\end{enumerate}
The first two subsets, \textit{R} and \textit{U} overlap, of course, but \textit{U} and \textit{O} do not. 
 Software other than PRAIA was used to reduce $D_{MPC}$ and its subsets.
We can now calculate the differences 
in the positions of Apophis between the subsets $D_{MPC}$ and $D_{PRAIA}$. 
The results are presented in Table~\ref{T:statdoublon}. 
Subset \textit{R} should allow us to study the effect of the reduction if it is greater than the UCAC4 intrinsic random errors.

As is shown in first line in Table~\ref{T:statdoublon}, 
the choice of the reduction software does not seem to influence
the astrometric results, provided the same catalog (UCAC4) is used.
The corresponding root mean square (RMS) values in right ascension and declination are a mix of the measurement uncertainties of Apophis itself and the errors contained in the UCAC4 catalog, both random and systematic.

Yet, they are well within the expected range of UCAC4 catalog errors 15-100~mas \citep{Zacharias12}.
Assuming the unknown systematic errors are randomly distributed, we conclude that intrinsic errors in the astrometric software suites are not visible here as a result of both the  measurement uncertainties of Apophis' position and the UCAC4 intrinsic catalog errors.

We proceed to study the effect of using different catalogs of the UCAC series in the astrometric
reduction process.
If systematic differences in positions of stars 
within the UCAC catalog series are of the same direction and magnitude, 
there should be no perceptible difference in exchanging individual UCAC catalogs during the reduction. 
The paired  Student's t-test applied to the subset \textit{U} 
rejects this hypothesis for right ascension at a $0.1\%$ confidence level or even smaller. 
Although there is no discernible effect in declination for this subset. 
Performing basic statistics on the Gaia-FUN-SSO campaign observations alone 
 allows us to conclude that there are differences  
between individual UCAC catalogs at least in right ascension. 
This result is supported by the recent catalog debiasing analysis in \citet{farnocchia-2015}. 
The RMS values of the position 
differences regarding the subset \textit{U} and the corresponding PRAIA reduced data 
are below $0.1\arcsec$. Therefore, they are still in perfect 
agreement with corresponding theoretical predictions for the expected RMS 
values in the UCAC catalogs series \citep{Zacharias04, Zacharias10, Zacharias12}, suggesting 
 that the dominant part of the "reduction" RMS in 
Table~\ref{T:statdoublon} consists of UCAC random errors.

In constrast, astrometric reductions with different tools \textit{\textup{and}} catalogs (\textit{O}) 
show a pronounced increase in RMS values, especially in right ascension. 
The RMS values of \textit{O} are 1.5-2 times greater than the corresponding values in \textit{U}.  

Assuming that the errors originating from both reduction ($\sigma_{red}$) and the catalog ($\sigma_{cat}$) are mutually independent, it is possible to deduce the effect 
of using different catalogs in astrometric reduction, i.e., 
\begin{equation}
\sigma_{cat}^2=\sigma_{cat+red}^2-\sigma_{red}^2.
\label {eq:sigma}
\end{equation}

Then, $\sigma_{cat, \alpha}=0.18\arcsec$ and $\sigma_{cat, \delta}=0.09\arcsec$ in our case. 
These kinds of errors are noticeable in the results of our astrometric reduction. They are mostly due to 
the contribution of the USNO-B1.0 catalog, which has declared errors around $0.2\arcsec$ \citep{Monet03} as the USNO-B1.0 catalog has a large relative part, at least 73\%, in the reduction of the positions present in subset \textit{O}. 
The additional shift of the mean values based on data set \textit{O} is almost of the same magnitude as the UCAC4 catalog errors. 
Again, the USNO-B1.0 catalog is the likely culprit, causing significant offset of the data mean values.
In addition, we must recall that USNO-B1.0 provides poor proper motions with an epoch of reference that is too far, around 1980-90  for its mean positions, which can be a source of errors.
Therefore, we recommend against using this catalog  for astrometric reduction anymore and to substitute it with the UCAC4 catalog. 

The conclusion of preceding analysis is clear: the dominant source of uncertainties in contemporary asteroid astrometry are (still) catalogs for the cases when we can neglect instrumental errors.
Unfortunately, the catalog USNO-B1.0 continues to be popular in astrometric reduction of asteroids and comets despite the fact that  better catalogs are available. 
We expect that once the Gaia astrometric catalog is available, catalog errors should no longer be the limiting factor for ground based astrometry \citep{perryman-2001}.
Until then, we strongly recommend migrating to more accurate reference catalogs containing 
not only accurate star positions but also reliable stellar proper motions (e.g., UCAC4) \citep[]{farnocchia-2015}.

\begin{table*}[]
\caption{Basic statistics (mean values and RMS) on differences in astrometric positions of Apophis between the observation sets $D_{MPC}$ and $D_{PRAIA}$ as defined in section~\ref{sec:homogeneous}.
The values are given in arcsec. 
Observations that have been reduced with different tools, but using the same catalog (UCAC4)
are denoted by \textit{R}. \textit{U} represents the astrometric results derived  using different catalogs of the UCAC series.
\textit{O} denotes the set of observations that were reduced using various software packages and catalogs other than UCAC. See text for details.}
\label{T:statdoublon}
\centering
\begin{tabular}{ccccc}
\hline\hline
 Samples & $\langle {(\alpha_{MPC}-\alpha_{PRAIA})\cos \delta_{PRAIA}}\rangle$ & $\sigma_{\alpha} \cos \delta$ &  $\langle {\delta_{MPC}-\delta_{PRAIA}} \rangle$ & $\sigma_\delta$ \\
 \hline
 \smallskip
 R  & $+0.005 \pm 0.005$ & $0.059$ &  $-0.000\pm 0.006$ & $0.079$\\
 U  & $-0.028 \pm 0.005$ & $0.096$ &  $-0.007\pm 0.005$ & $0.098$\\
 O & $+0.069 \pm 0.016$ & $0.205$ &  $+0.022\pm 0.010$ &  $0.132$\\
\hline
\end{tabular}
\end{table*}

 \section{\addb{Alerts and recovery process}}
 \label{sec:alerts}

The main purpose of Gaia-FUN-SSO is to recover and track new objects discovered by the Gaia satellite. 
A strategy for recovery and follow up has been investigated in \cite{bancelin12}, which combines space and
ground-based data  to optimize the network's performance. Using statistical tools and nonlinear orbit propagation (Monte Carlo
technique), those authors analyzed the evolution of the size of the initial ($\alpha$,$\delta$) distribution, i.e., how the uncertainties in the
astrometric positions change with time after the detection of an object by the Gaia satellite. 
In this respect, we would like to know
whether the present method used for data reduction
has a significant impact on follow up within the network.
Following an approach similar to that of \cite{bancelin12}, we aim to assess how far the
predicted position can drift from the real one in a given amount of time.
We consider a hypothetical discovery of an asteroid during the Gaia-FUN-SSO campaign. We use the observational data on Apophis,
but we assume its orbit was previously unknown.
Furthermore, we assume that the hypothetical discovery has happened on the first night recorded in the duplicated measurements
D$_{\scriptscriptstyle PRAIA}$ and D$_{\scriptscriptstyle MPC}$. We include the three post reduction outliers again into our data set
to have a more realistic sample; see section \ref{sec:homogeneous}. Hence, D$_{\scriptscriptstyle PRAIA}$ and D$_{\scriptscriptstyle MPC}$ encompass 629 observations. 
However, we only use the data from the first night to determine the orbit and orbital element covariance matrix of the new object.
We applied individual statistical
weights according to the respective observatory code. Those weights can be found in \citep{chesley-2010}. 

We then propagated the orbit
solutions and uncertainties obtained from both sets up to six days after
the discovery using the OrbFit open source software package\footnote{\url{http://adams.dm.unipi.it/orbfit/}}.

\begin{figure}[h!]
 \centerline{
\includegraphics[angle=0,width = \columnwidth]{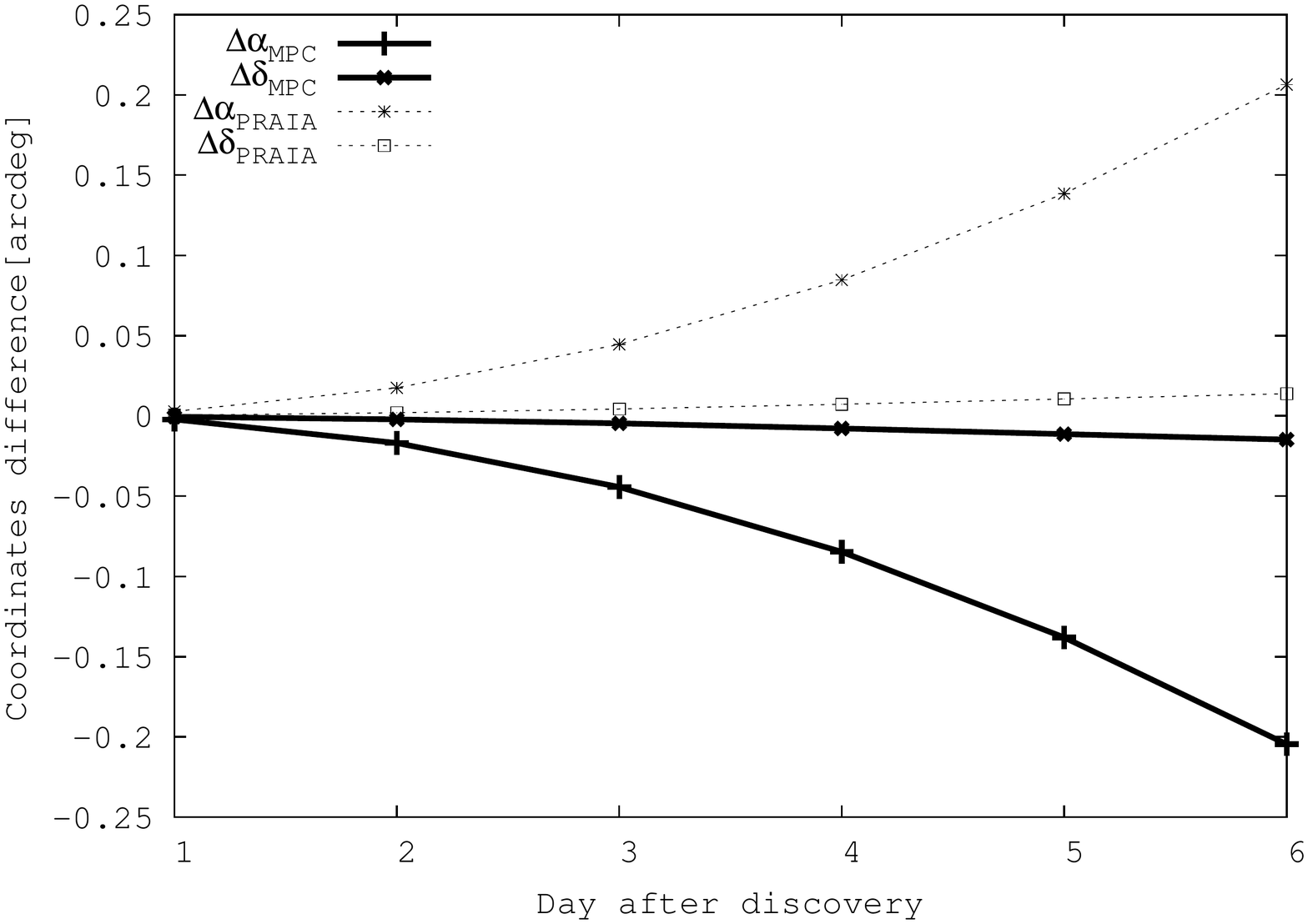}
}
\caption{Impact of different reduction pipelines on the recovery of newly found NEOs.
The graph shows the time evolution of the coordinate differences ($\Delta \alpha$, $\Delta \delta$)$_{\scriptscriptstyle MPC}$ and
($\Delta \alpha$, $\Delta \delta$)$_{\scriptscriptstyle PRAIA}$ between orbit solutions derived 
from different data sets with respect to the nominal solution. The nominal orbit of Apophis has been derived using all available data (optical and radar).
The MPC or PRAIA solutions are derived from fitting orbits to observations performed during the first night of the duplicated data sets 
$D_{PRAIA}$ and $D_{MPC}$.}
\label{F:evolution}
\end{figure}

One week after the discovery, the coordinate differences $\Delta \alpha$ and $\Delta\delta$ between D$_{\scriptscriptstyle PRAIA}$, D$_{\scriptscriptstyle MPC}$
and the "true" position of Apophis are evaluated. The latter has been generated using all available observations form 2004-2014 (optical
and radar). Figure \ref{F:evolution} shows how the differences in astrometric coordinates evolve for both sets of measurements during the six days following the discovery. The
opposing orientation of the ($\Delta \alpha$, $\Delta \delta$)$_{\scriptscriptstyle MPC}$ and ($\Delta \alpha$, $\Delta
\delta$)$_{\scriptscriptstyle PRAIA}$ curves is due to the different preliminary orbital elements found using D$_{\scriptscriptstyle PRAIA}$ and D$_{\scriptscriptstyle MPC}$. 
One can see that ($\Delta \alpha$, $\Delta \delta$)$_{\scriptscriptstyle MPC}$ and
($\Delta \alpha$, $\Delta \delta$)$_{\scriptscriptstyle PRAIA}$ are of the same order of magnitude. Consequently,
the method of data reduction is unlikely to have a significant impact on the recovery process
within the network.


 \subsection{Position uncertainty propagation for new discoveries}
\begin{figure}[h!]
 \centerline{
\includegraphics[angle=0,width = \columnwidth]{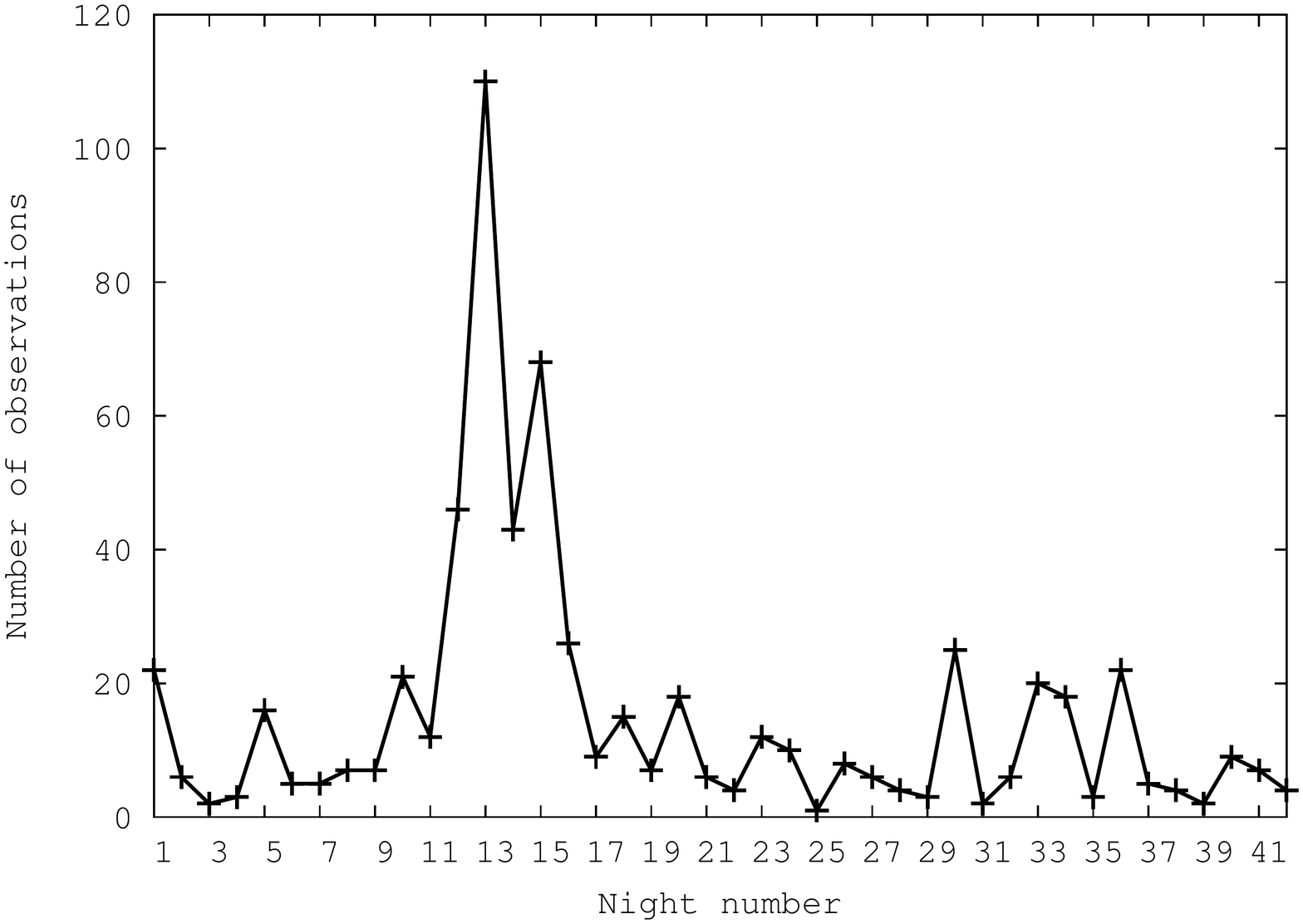}
}
\caption{Number of observations per observed night for the duplicated measurements.}
\label{F:obs_night}
\end{figure}

Now we see how the position uncertainty evolves when more observations become available during the nights following an asteroid's discovery. 
To this end, we only used the duplicated set of measurements D$_{\scriptscriptstyle PRAIA}$ and D$_{\scriptscriptstyle MPC}$. They
contain 42 nights encompassing 629 observations; see Fig. \ref{F:obs_night}. 
Please note that observation nights are not necessarily consecutive. Sometimes days may pass without any observation being performed.

\begin{figure}[h!]
 \centerline{
\includegraphics[angle=0,width = \columnwidth]{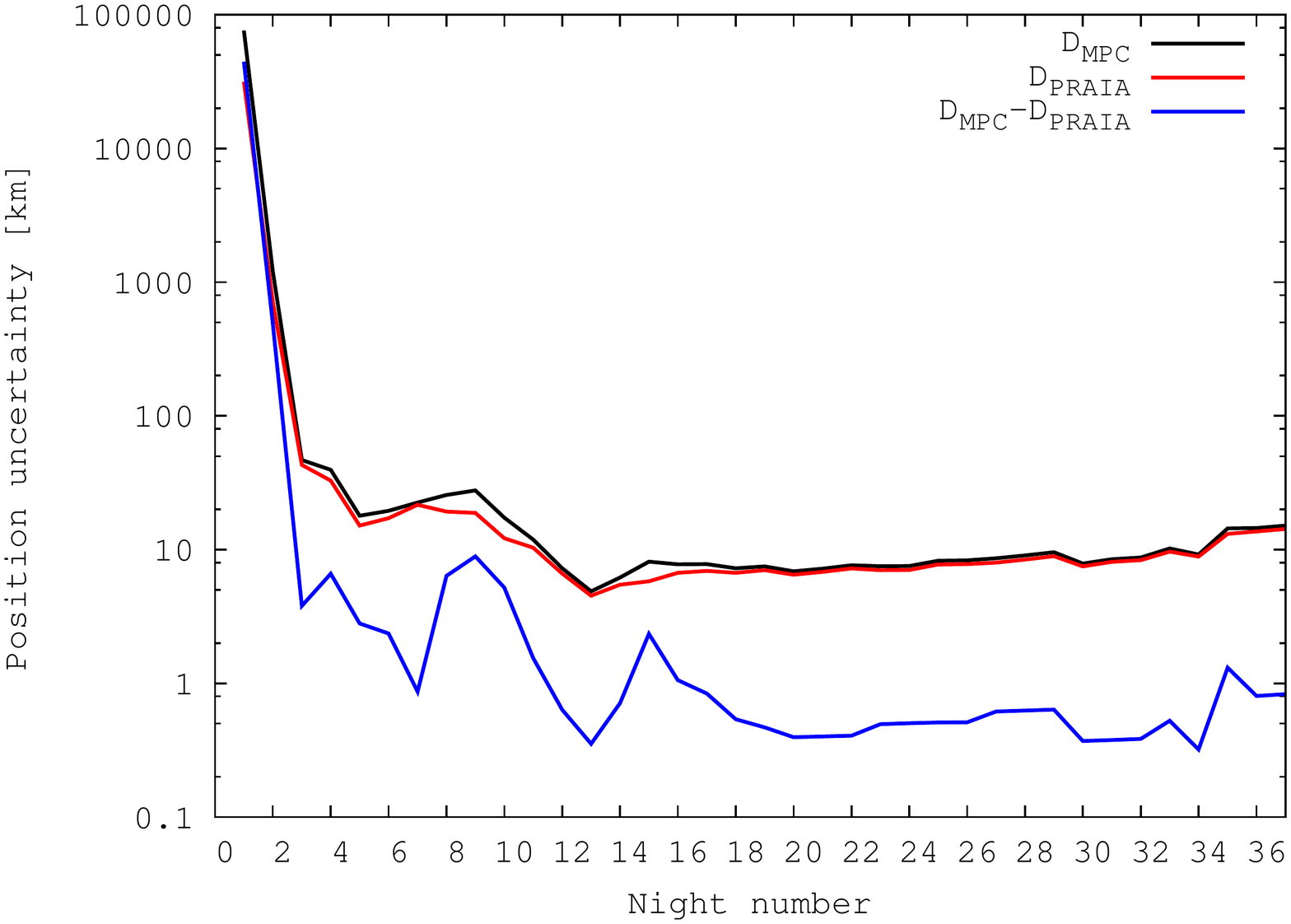}
}
\caption{Geocentric position uncertainty evolution as a function of the number of observation nights for the duplicated measurement
sets
D$_{\scriptscriptstyle PRAIA}$ and D$_{\scriptscriptstyle MPC}$. The positioning uncertainty is updated successively as soon as new data
from another observation night becomes available. The difference between the two data sets is also indicated.}
\label{F:uncertainty}
\end{figure}

As we assume the asteroid is newly discovered, a preliminary orbit determination is conducted after the first night. 
A sequential orbital improvement is performed whenever new observations become available and uncertainties in the geocentric distance of the "newly discovered" Apophis are calculated. 
This allows us to compare the impact of the reduction pipeline on the uncertainty evolution of a newly
found object. Figure~\ref{F:uncertainty} shows that the astrometric uncertainties are large for both D$_{\scriptscriptstyle MPC}$ and D$_{\scriptscriptstyle PRAIA}$
data after the first night (discovery night). However, the uncertainties are slightly smaller when the orbit is computed from D$_{\scriptscriptstyle PRAIA}$ data. 
This is the result of  better precision of the corresponding set of observations; see section \ref{sec:homogeneous}. 
 The difference between the uncertainty computed with MPC and PRAIA data (denoted as 
D$_{\scriptscriptstyle MPC}$- D$_{\scriptscriptstyle PRAIA}$ in Fig.~\ref{F:uncertainty}) drops permanently below 10 km only after the $10^{th}$ observation night.
Since the first and $10^{th}$ observation night span an arc of 26 days, encompassing 96 observations, there is a real advantage in our scheme of reduction regarding the position uncertainty propagation of follow-up campaigns.

\section{Position uncertainty propagation for Apophis}
\label{sec:bplane}

In this section, we  investigate the influence our scheme for reducing astrometric data has on propagated uncertainties in the 2029 b-plane distance \citep{valsecchi-2003} between Apophis and the Earth. 
\subsection{The target plane (b-plane)}
The target plane (b-plane) is a useful concept for describing close approaches between asteroids and planets. 
The b-plane is, in fact, a generalization of the impact parameter used to characterize two-body scattering processes. 
Passing through the Earth's center, the b-plane is perpendicular to the geocentric velocity of the asteroid directed along the incoming asymptote of its hyperbolic orbit with respect to the Earth. 
Any location on the target plane can be specified using two geocentric coordinates
($\xi,\,\zeta$). If the uncertainties in the asteroid's orbital elements are modeled by multivariate Gaussians centered on the nominal orbit, 
the projection of the six-dimensional uncertainty space on the b-plane resembles an ellipse centered around the point of 
intersection of the incoming asymptote of the nominal orbit with the b-plane
($\xi_{\scriptscriptstyle 0},\,\zeta_{\scriptscriptstyle 0}$). The semimajor and semiminor axes of this projected ellipse are equal to
$3\sigma_{\scriptscriptstyle \zeta}$
and $3\sigma_{\scriptscriptstyle \xi}$ respectively. Accordingly, the distance of the closest approach is equal
to:$\sqrt{\xi_{\scriptscriptstyle 0}^{\scriptscriptstyle 2} + \zeta_{\scriptscriptstyle 0}^{\scriptscriptstyle 2}}$. The b-plane is defined such that 
the uncertainty in the asteroid-to-Earth distance lies in the $\zeta$ component and is approximately equal to
$\sigma_{\scriptscriptstyle \zeta}$.

\subsection{Orbit fitting and uncertainty propagation}
We  proceed to study whether orbits and initial uncertainties constructed from different sets of observations can cause a 
significant change in the propagated uncertainties of Apophis' orbit.
One obstacle we  encountered in this process is the fact that Apophis is most likely in a tumbling rotational state, as pointed out by \cite{pravec-2014}.
As no dynamical model is currently available that describes the Yarkovsky effect for tumbling asteroids in a satisfying way, 
there is no point in trying to achieve a highly accurate prediction for Apophis' future orbit without a detailed analysis of this problem; this is beyond the scope of this work. Instead, we  decided to use a simplified dynamical model, including 
gravitational and relativistic interactions with the sun and the planets as well as possible severe perturbations from the asteroid belt, to compare the relative changes in the orbit uncertainties caused by  
different sets of observations. The process  works as follows.
After an initial orbit determination, an orbit adjustment based on a differential correction is performed. This results in the uncertainties in the asteroids orbit in form of 
an orbital element covariance matrix.
 The resulting uncertainties can then be propagated to the date of the close encounter of Apophis and the Earth on 13/04/2029.
Then, the propagated uncertainty hyper-volume is projected onto the 2029 b-plane and its long axis was used to
indicate the 1$\sigma$ uncertainty value.
Similar to section \ref{sec:alerts}, we used the OrbFit package  for orbit determination and propagation, and we applied  statistical weights
from \citet{chesley-2010}. Because of the high offset of the observatory B04 with respect to the JPL ephemerides discovered in Table \ref{T:reduction}, a  lower weight  was given to improve the global residuals. 

\add{
A quick first check can be performed using the duplicated measurement sets D$_{\scriptscriptstyle PRAIA}$ and D$_{\scriptscriptstyle MPC}$ as the only source for initial orbit generation. 
The timespan covered by the duplicated observations is only four months, however. Thus, the propagated uncertainty to the 2029 b-plane is very high for both sets, but
the propagated nominal solution obtained with D$_{\scriptscriptstyle PRAIA}$ improves the 1$\sigma_{\scriptscriptstyle
\zeta}$ uncertainty obtained with D$_{\scriptscriptstyle MPC}$ by $\sim$14\%, which is non-negligible for the impact probability assessment
with short arc data. This short test confirmed that a set of homogenous data (D$_{\scriptscriptstyle PRAIA}$) reduced with
the same  precise astrometry software  and a relevant  reference catalog results in improved orbit uncertainties.}

\subsection{Impact of Gaia-FUN-SSO observations on orbit uncertainties}
Our aim is to investigate whether the data produced during the Gaia-FUN-SSO campaign can impact orbital solutions and b-plane
uncertainties through the example of Apophis. To this end, we compare orbits and uncertainties derived from five observational data sets with respect 
to their influence on the close encounter b-plane in 2029. 
The existing radar observations performed between 2005 and 2013 were included in each set
in order not to overestimate the impact of the new optical data. The radar observations are taken from the corresponding JPL
website\footnote{\url{http://ssd.jpl.nasa.gov/?grp=all&fmt=html&radar=}}.
\add{} The first set (S$_{\scriptscriptstyle 1}$)  includes all the optical observations of Apophis as presented in the MPC database. 
The set S$_{\scriptscriptstyle 2}$ contains the same number of observations as
S$_{\scriptscriptstyle 1}$, but the old duplicate measurements  D$_{\scriptscriptstyle MPC}$ have been replaced by the newly reduced set D$_{\scriptscriptstyle
PRAIA}$. In other words, the duplicated observations reduced with different catalogs have been exchanged with those that
were reduced in a homogeneous manner. 

The set S$_{\scriptscriptstyle 3}$  left the data submitted to the MPC unchanged. We only added the previously unpublished optical observations  so that
S$_{\scriptscriptstyle 3}$ is composed of S$_{\scriptscriptstyle 1}$ and S$_{\scriptscriptstyle NEW}$. 
The set S$_{\scriptscriptstyle 4}$  combines old measurements with all available newly reduced data. It is, thus, composed of set S$_{\scriptscriptstyle NEW}$ and S$_{\scriptscriptstyle 2}$,
where S$_{\scriptscriptstyle NEW}=2103$ previously unpublished astrometric observations. This means we replaced  old duplicate measurements  by D$_{\scriptscriptstyle
PRAIA}$ in this set.
Both S$_{\scriptscriptstyle 3}$ and S$_{\scriptscriptstyle 4}$ contain the same number of observations. 
Finally, we are also interested in the
orbital accuracy that can be achieved  using the 2732 Gaia-FUN-SSO observations reduced by PRAIA exclusively. This set
is called S$_{\scriptscriptstyle 5}$.  We  summarize the observational sets
definition as follows:

\begin{itemize}
 \item S$_{\scriptscriptstyle 1}$ = [2004-2014]$_{\scriptscriptstyle MPC}$ + radar
 \item S$_{\scriptscriptstyle 2}$ = [2004-2014]$_{\scriptscriptstyle MPC}$ - D$_{\scriptscriptstyle MPC}$
+ D$_{\scriptscriptstyle PRAIA}$ + radar
 \item S$_{\scriptscriptstyle 3}$ = S$_{\scriptscriptstyle 1}$ + S$_{\scriptscriptstyle NEW}$
 \item S$_{\scriptscriptstyle 4}$ = S$_{\scriptscriptstyle 2}$ + S$_{\scriptscriptstyle NEW}$
 \item S$_{\scriptscriptstyle 5}$ = S$_{\scriptscriptstyle NEW}$ + D$_{\scriptscriptstyle PRAIA}$ + radar,
 
\end{itemize}
where [2004-2014]$_{\scriptscriptstyle MPC}$ refers to the 4138 optical data as present in the MPC database.

We propagated each nominal orbit resulting from the individual sets of observations along with its covariances up to 2029 where we evaluated the position uncertainties 
projected onto the b-plane. Table~\ref{T:b-plane} summarizes the quality of the orbital fit and the 2029
b-plane uncertainty resulting from the orbit propagation. The table contains the uncertainty propagation results, as well as the quality test of the orbital fit, i.e., the
reduced $\chi_{red}^2$ for optical and radar data, as follows: 

\begin{eqnarray}
\label{eq:chi2}
\chi^2_{red}&=&\frac{1}{2N+n-p-1}\left(\sum_{i=1}^{N} \left(\frac{(O-C)_\alpha}{\sigma_\alpha}\right)_i^2 \right.\\\nonumber
      &+&\left. \sum_{j=1}^{N} \left(\frac{(O-C)_\delta}{\sigma_\delta}\right)_j^2+\sum_{l=1}^{n} \left(\frac{(O-C)_r}{\sigma_r}\right)_l^2\right), 
\end{eqnarray}

where $N$ the number of optical observations and $n$ is the number of radar observations and $p=6$ is the number of fit parameters.
The observed values ($O$) are the astrometric measurements in the respective sets $S_1$-$S_5$, and the calculated values ($C$) are from
the corresponding orbits. To be compatible with the orbit fitting and propagation, 
the variances in equation (\ref{eq:chi2}) were derived from the weighting schemes proposed in \cite{chesley-2010}.
We fit the radar and optical observations together. However,  to demonstrate the effect of the differently reduced observations
on the radar and optical fits, we present reduced $\chi^2$ values for the radar and optical measurements in separate columns in Table~\ref{T:b-plane}.

\begin{table}[h!]
 \begin{center}
  \caption{Orbital accuracy information, fit residuals and b-plane uncertainty, computed with different sets of
observations. See text for the definition of each set. We also computed the difference in b-plane distance
$\Delta_{\scriptscriptstyle i}$ for each set with respect to the distance $\Delta_{\scriptscriptstyle 1}$ obtained from
S$_{\scriptscriptstyle 1}$
  }
  \label{T:b-plane}
  \begin{tabular}{|c|c|c|c|c|c|}
   \hline
   \hline
   & $\chi^2_{\scriptscriptstyle {red}}$   & $\chi^2_{\scriptscriptstyle {opt}}$&$\chi^2_{\scriptscriptstyle {rad}}$ &  $\sigma_{\zeta}$  (km)&
$\Delta_{\scriptscriptstyle i}$ -$\Delta_{\scriptscriptstyle 1} $ (km)\\
  \hline

S$_{\scriptscriptstyle 1}$ &0.227 & 0.227& 0.434 & 2.99 & 0\cr
\hline  
S$_{\scriptscriptstyle 2}$ &0.224 &0.224& 0.426& 2.94 & 0\cr
\hline
 S$_{\scriptscriptstyle 3}$ &0.157 &0.157& 0.175 & 2.45 & 1.5\cr
\hline
S$_{\scriptscriptstyle 4}$ & 0.155&0.155  & 0.174 & 2.43 & 1.5\cr
\hline
 S$_{\scriptscriptstyle 5}$ & 0.021 &0.021& 0.095 & 3.24 & 3\cr

\hline
  \end{tabular}
 \end{center}
\end{table}


Clearly, the contribution of optical measurements is dominating the $\chi^2_{red}$ because of the
large ratio $N/n$. The values of $\chi^2_{red}$ are smaller than unity indicating 
that the chosen variances in equation (\ref{eq:chi2}),  which were taken from \cite{chesley-2010}, are not well suited to describe the distribution of the residuals. In fact, they are too large. This may be because the historical performance on the
observatory sites is based on observations of NEOs that are much fainter than Apophis
with correspondingly lower signal-to-noise ratios. A proper weighting should  bring ${ \chi^2_{ red}}$  closer to one.

Nevertheless, the  presented data suggest that the sets containing D$_{\scriptscriptstyle PRAIA}$ instead of D$_{\scriptscriptstyle MPC}$ 
result in smaller uncertainties in Apophis' positions in the 2029 b-plane. Indeed,
even for a well-known orbit (with a 10-years arc data length),
both optical and radar $\chi^2$ values show better results when
 D$_{\scriptscriptstyle PRAIA}$ measurements are used. 
Hence, we speculate that current orbit solutions of NEAs can be improved using  data  issued from, at first, stellar astrometry based on an accurate catalog and, to a lesser extent, homogeneous reduction with a reliable software.
Furthermore, the use of this kind of data can also result in smaller uncertainties in the b-plane coordinates of PHAs,
as was shown for the 2029-b-plane of Apophis, cf. ($S_1$, $S_2$) and ($S_3$, $S_4$).
If the new measurements provided with this article are taken into account, the b-plane uncertainty can most likely be reduced considerably,
cf. ($S_1$, $S_3$) and ($S_2$, $S_4$).
\add{Moreover,} we see that Gaia-FUN-SSO and radar data \add{(S$_{\scriptscriptstyle 5}$)} suffice to produce b-plane
uncertainty values that are very close
to those sets containing all available observations. 

\section{Imprecise observer location coordinates}
\label{sec:geoposition}
Since its discovery in June 2004, Apophis has had several close encounters with the Earth. Two times the close encounter distances were less than 0.1~au. 
If objects pass by the Earth at such close range, parallax effects in topocentric directions have to be very accurately computed in the subsequent analysis, as
parallactic shifts can reach more than one degree for objects coming within one lunar distance to the center of the Earth.
A consequence of the increased parallax, uncertainties in an observer's location can become a non-neglibile source of errors in astrometric results during 
close encounters of NEOs with the Earth. 
Having access to the precise location at which any given observation has been performed becomes paramount under such circumstances.  
In Appendix A we provide a simple framework to quantify requirements on the precision of the geolocation of an observer as a function 
of the topocentric distance to the observational target.

As stated earlier, Apophis had two close encounters with the Earth since its 2004 discovery: one on December~21, 2004 at 0.09638~au and one on January~09, 2013 at 0.09666~au. 
If we require the positioning error due to the parallax effect ($\epsilon_{plx}$) to be negligible, e.g., $\epsilon_{plx}<0.005\arcsec$, 
we find that the geolocation of the observers should be known to within 350~m or $0.003\degr$ in angular coordinates. 
After checking the positions of each of the observational stations of the Gaia-FUN-SSO, which participated in the Apophis campaign, we compared 
them with the 
geolocation data of the observers kept at the MPC\footnote{http://www.minorplanetcenter.net/iau/lists/ObsCodes.html}.
We found that there are differences greater than 350~m in the geolocation of a few telescopes that participated in the GAIA-FUN-SSO campaign. In fact, 
one observatory had geopositioning errors as large as $7.6~$km. As shown in 
Table~\ref{T:staterror}, the correction of the geocentric coordinates decreased mean values of the total $(O-C)$ of astrometric positions, 
and, to a smaller extent, the associated RMS values.
Here, the $O$ values denote that the observations reduced with PRAIA and UCAC4, the $C$ values were the predictions by the JPL Horizon ephemerides service, and the 
averages $\langle O-C \rangle$ are calculated using the 2732 measurements generated during the Gaia-FUN-SSO campaign.

Furthermore, there is only one geolocation associated with a single MPC code. Yet, there can be several telescopes 
at one site performing observations of asteroids, which is not resolved by the MPC format within one observatory code. If the positions between two telescopes at a known observatory differ by more than $350~m$, 
this information should be made public,
as it  influences the site's astrometric data quality and statistics.
Another difficulty arises when telescopes are refurbished, replaced, or relocated within the same observatory, since there is no chronology of instruments publicly available at the MPC.

We strongly suggest  following the advice of the MPC and checking the locations of observation sites, since the systematic error associated with the inaccuracy of the observers' position propagates into the orbit fitting of asteroids.
A guide on how to best accomplish an update of the positions of a given instrument can be found on the MPC web page\footnote{http://www.minorplanetcenter.net/iau/info/ObservatoryCodes.html}.

\begin{table}[htb!]
\caption{Influence of errors in geolocation on the (O-C) of Apophis in units of arcsec.}
\label{T:staterror}
\centering
\begin{tabular}{c r r}
\hline\hline
 585 Observatory & MPC geolocation & Updated geolocation\\
 \hline
 \smallskip
 $\langle {(O-C)_\alpha \cos \delta}\rangle$ & $-0.057$ & $-0.035$\\
 $\sigma_{\alpha} \cos \delta$ & $0.081$ & $0.077$\\
 \smallskip
 $\langle {(O-C)_\delta}\rangle$ & $+0.042$ & $+0.004$\\
 $\sigma_\delta$ & $0.075$ & $0.074$\\
 \hline\hline
 O44 Observatory & MPC geolocation & Updated geolocation\\
 \hline
 \smallskip
 $\langle {(O-C)_\alpha \cos \delta}\rangle$ & $-0.029$ & $-0.029$\\
 $\sigma_{\alpha} \cos \delta$ & $0.034$ & $0.034$\\
 \smallskip
 $\langle {(O-C)_\delta}\rangle$ & $+0.067$ & $+0.054$\\
 $\sigma_\delta$ & $0.048$ & $0.049$\\
\hline
\end{tabular}
\end{table}

\section{Rounding of astrometric data}
\label{sec:rounding}
The presented new astrometric measurements have not been rounded to the usual data format, as both MPC formats with and without extra precision are compatible with the MPC standard.

This section contains a brief discussion on the reasoning behind using extended precision for all data. 

Observations of stations with an MPC designation that show reasonably small residuals are published in the Minor Planet Supplement (MPS). 
The usual data format supported by the Minor Planet Center (MPC) allows observers to provide their 
astrometric measurements with a precision of five post comma digits in the fractional part of the day, two 
post comma digits in seconds of right ascension, and one post comma digit in declination. 
In principle, the MPC format allows for an increased  
precision (one more digit) in both, observation epoch, and position.
Depending on the sites' equipment and performance, observers may take advantage of the extended supply of post-comma digits
with their data. High precision observations are, of course, very valuable for orbit fitting procedures.
Estimating the accuracy of the supplied data, however, can be difficult. If it is uncertain whether the quality of the measurements is high enough
to warrant extended precision, results are often rounded to the closest value within the usual MPC format.
We consider this practice to be suboptimal for several reasons.
First, this procedure modifies the original data, hampering reproducibility. Second, accidentally rounding "good data" introduces an additional source of error. 
Third, keeping an extra digit does not cause any harm, since no additional modifications in the format are necessary.

\begin{figure} \resizebox{\hsize}{!}{\includegraphics{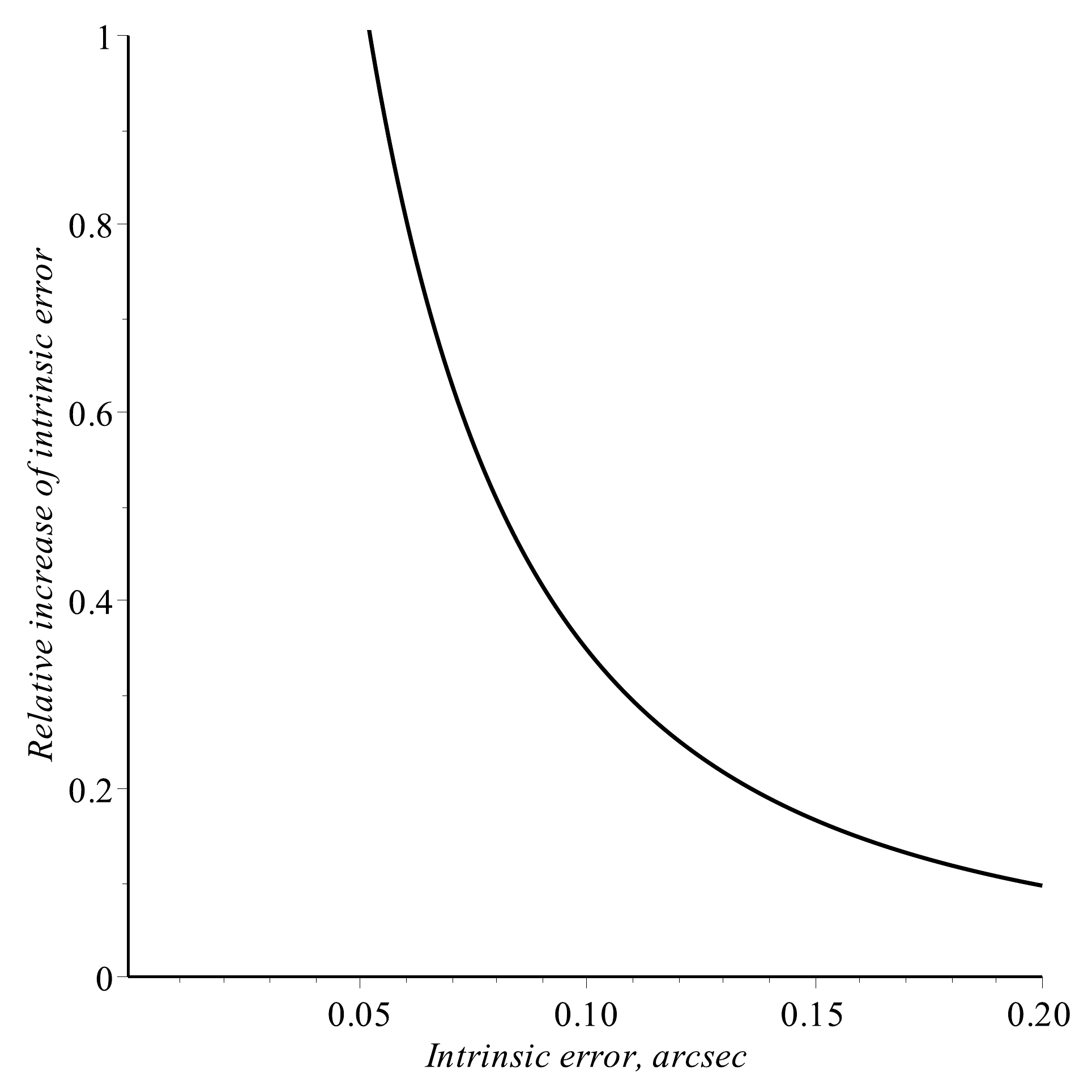}} \caption{Increase of the intrinsic observational error due to rounding in MPC format.}
\label{F:error}
\end{figure}

To illustrate those points, we  quantify the influence of rounding on the overall astrometric precision.
A proxy for the worst scenario, the maximum deviation of the position represented in the MPC standard format 
reads as follows: 
\begin{equation}
\sigma_{max}=\sqrt{(\sigma_{\alpha}^2+\rho_\alpha^2)\cos ^2 \delta + (\sigma_{\delta}^2+\rho_\delta^2)},
\end{equation} 
where $\sigma_{\alpha}$ and $\sigma_{\delta}$ are intrinsic uncertainties of the astrometric position. 
As stated previously, the maximum error contribution due to 
rounding is given by $\rho_\alpha=0.075\arcsec$, and $\rho_\delta=0.05\arcsec$.
In the case of observations close to the celestial equator and negligible position uncertainties ($\sigma_{\alpha}=\sigma_\delta=0$), 
the rounding of positions to the standard MPC format introduces a maximum error of $\sigma_{max}=0.09\arcsec$. 
Fig.~\ref{F:error} presents the maximum relative contribution of the rounding error to the intrinsic observation uncertainties. 
For an intrinsic observational error $\sigma_{obs}=\sqrt{\sigma_{\alpha}^2\cos ^2 \delta + \sigma_{\delta}^2}= 0.1\arcsec$, one  finds a relative 
contribution of the round-off errors of as much as $34\%$.  
The relative contribution of rounding errors is still visible ($10\%$) for intrinsic observational uncertainties of $0.2\arcsec$. 
Although the errors introduced by rounding are not excessive, they are unnecessary and should be avoided.
The conducted analysis assumes, of course, that the last digit of the measurement was significant. Hence, we conclude that there is no reason to round the original astrometric
positions, especially if the data were reduced with UCAC catalogs, which have uncertainties at the level of $0.1\arcsec$, and if the observational error is less than $0.3 \arcsec$, which keeps the rounding error contribution to less than 5\% of the total error budget.

We now discuss rounding errors in the observational epoch.
The maximum round-off error for the observational epochs in the standard MPC format is $0.432$s. This kind of an error shifts the position of the moving asteroid along its trajectory 
by 
\begin{equation}
d=\rho_t \sqrt{{v_{\alpha}}^2\cos^2_{\delta}+{v_{\delta}}^2},
\end{equation}
where $\rho_t=0.432$s denotes the round-off error of the usual MPC format and the apparent velocity components $v_{\alpha}$ and $v_\delta$ are given in $[\arcsec/s]$. 
This sort of error depends on the apparent velocity of the object with respect to the background stars in field of view.
The maximum apparent velocity of Apophis during its close approach in 2004 was 
$0.118\arcsec/s$ observed on December 20, 2004. If we round the time moments of observations, according to the standard format of the MPC,  
one introduces a maximum displacement, i.e., error of $d=0.118\arcsec/s\;\rho_t=0.051\arcsec$ along the track.
For the apparition in 2013, one can find a corresponding displacement error of $d=0.026\arcsec$ on January 16, 2013. 
One can see that the rounding errors due to imprecise timing seem negligible compared to the 
measurement positional errors. 
However, during its close approach so far on April 13, 2029 Apophis is bound to produce displacement errors ranging from $d=0.159\arcsec$
to $d=18.9\arcsec$. 
This example serves to underline the point that fast moving objects require special treatment. Fortunately, the MPC offers guidelines on this subject.\footnote{http://www.minorplanetcenter.net/iau/info/VideoAstrometry.pdf} 
If standard astrometry is to be performed, the enforcing of extended precision in the observation epoch 
is mandatory for these kinds of cases.

\section{Summary}
\label{sec:conclusions}
This Gaia Follow Up Network for Solar System Objects (Gaia-FUN-SSO) has been set
up to facilitate ground based retrieval of asteroids
discovered during the Gaia mission.
Unfortunately, the Gaia team had to overcome several technical difficulties since the begining of this mission \footnote{ see for example http://blogs.esa.int/gaia/}. 
Therefore, the triggering of solar system alerts could not yet be tested.  The verification phase of this process has been initiated only late in 2014. Nevertheless,  the Gaia-FUN-SSO network carried out several observing campaigns  to train coordination. 
In particular, an astrometric observation campaign was launched  
during the latest period of observability of the PHA (99~942) Apophis in 2012-2013.

To test the network's coordination and performance, an astrometric observation campaign was launched  
during the latest period of observability of the PHA (99~942) Apophis in 2012-2013. 
A large amount of astrometric data was collected and processed in a homogeneous fashion using the PRAIA reduction software and the UCAC4 catalog data. 
The resulting 2732 precise astrometric
measurements recorded by 19 observatories worldwide are now available for community use. 

We have, furthermore, taken advantage of the fact that 629 measurements performed by the observers had already been sent to the Minor Planet
Center (MPC). As is common practice, different catalogs and different software were
used during the individual data reduction.
Since these data were reanalyzed using PRAIA and UCAC4 as well, this experiment has provided us with an opportunity to
test the impact of   proceeding with this approach on the accuracy of real near-Earth object astrometry.
We could show that there is a significant difference in the quality of the resulting measurements, which is mostly due to catalog biases.
Our reduction of observations resulted in a decrease of a factor of two in RMS uncertainties 
for the astrometric positions basically bringing them down to the level of UCAC4 catalog errors.
 
The choice of the astrometric analysis pipeline does not seem to have a large impact on the recovery process of new objects when
their observational data arcs span less than one night. However,
improved astrometric positions directly translate into a greater reduction of NEO position uncertainties during follow-up campaigns. 
Had Apophis been discovered during the Gaia-FUN-SSO campaign, a consistent data reduction with state-of-the-art software and catalogs
would have had a clear impact on the subsequent orbit predictions.

Using a simplified dynamical model, we have found that the new 2103 astrometric measurements presented in this paper are likely to have a 
significant impact on the position uncertainty of Apophis during its close encounter with the Earth in 2029. 

To further increase the astrometric precision of Gaia-FUN-SSO, we  identified the necessary requirements on the accuracy of geolocations for the
participating telescopes. In some cases, we  discovered severe discrepancies between the actual observer positions and those listed in the MPC database. 
Taking the opportunity to demonstrate the impact of
imprecise geopositioning on the resulting astrometric measurements, we urge 
observers to follow the MPC recommendation to check and update information on their sites.

Finally, we  discussed the impact of rounding of astrometric positions and observational epochs to the standard MPC format. Given the possibility of 
displaying data in an extended format, we see no advantage in rounding to the standard format, even if the precision is less than the last available digit. 
On the contrary, we have demonstrated that rounding can lead to additional errors that are best avoided. We suggest keeping an extra digit in the position each time the standard error is less than $0.3\arcsec$ and an extra digit in the observational epoch whenever this 
is supported by the hardware at hand.


\begin{acknowledgements}
       The authors are grateful to several engineers and technicians who punctually participated in this work.  
       W. Thuillot, S. Eggl and D. Hestroffer acknowledge the support of European Union Seventh Framework Program (FP7/2007-2013) under grant agreement no. 282703 (NEOShield project). 
       A. Ivantsov is grateful for the support by Mairie de Paris: Research in Paris 2012.
       J. Desmars was supported by CNPq grant 161605/2012-5. 
       M.Assafin acknowledges CNPq grants 473002/2013-2, 482080/2009-4 and 308721/2011-0, FAPERJ grant 111.488/2013. 
       We  acknowledge CAPES/COFECUB programme ASTRODYN/Te-791-13 for the Brazil-France cooperation. 
      The work performed at the RTT150/ TUBITAK National Observatory in Turkey was funded by the subsidy allocated to Kazan Federal University for the state assignment in the sphere of scientific activities. 
       Q.Y. Peng acknowledges financial support from the National Science Foundation of China with the Grant Nos (11273014 and C1431227). 
       This work is based in part on observations made at Observatoire de Haute Provence (CNRS), France.
       The authors are also grateful to D. Farnocchia (JPL) for fruitful discussions.

\end{acknowledgements}


\bibliographystyle{aa}
\bibliography{apophis}

\appendix
\section{Requirements for geolocation precision}
\begin{figure} \resizebox{\hsize}{!}{\includegraphics[angle=0,width = \columnwidth]{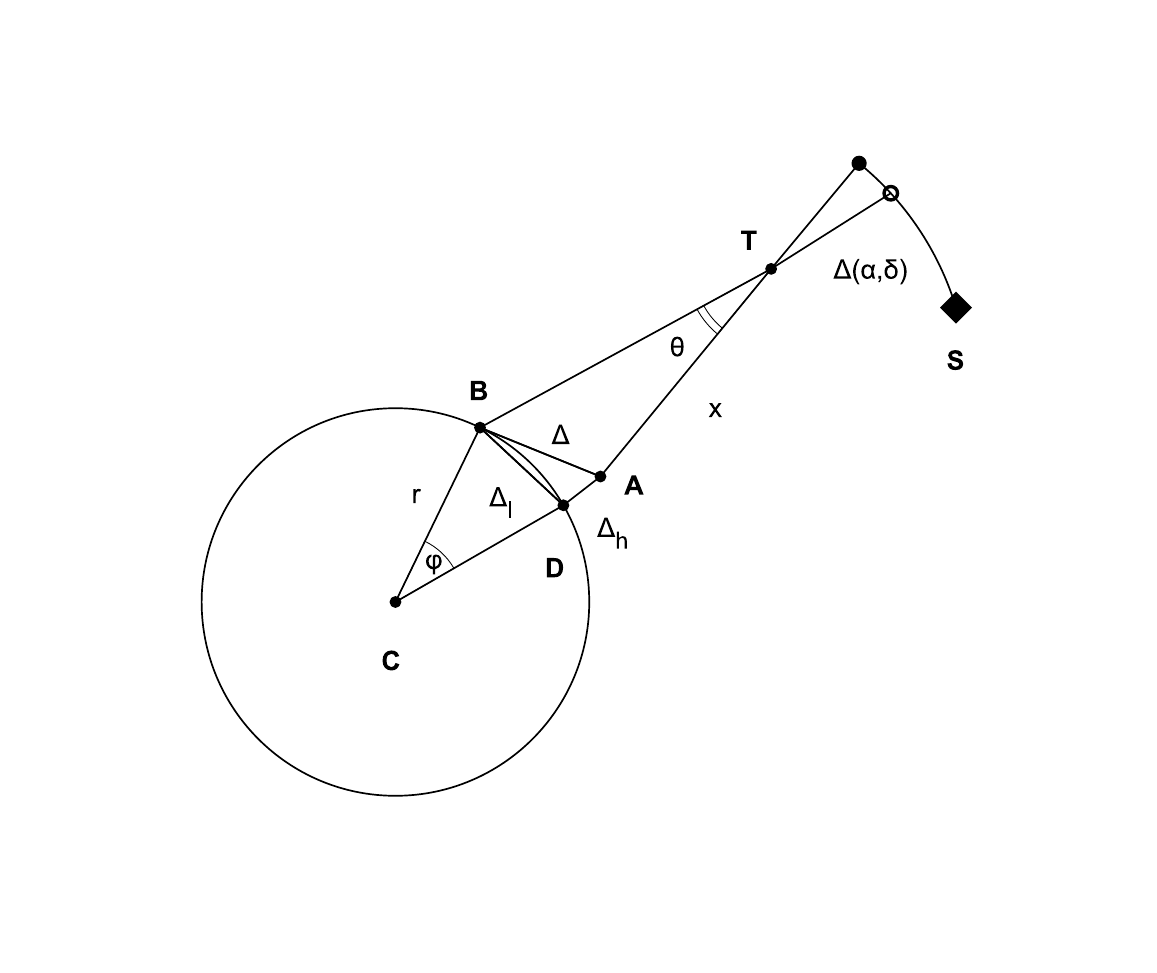}} \caption{Error in parallax reduction due to uncertainties in the position of observer. In the case of a small angle $\phi$, one can use ${\Delta}^2={\Delta_l}^2+{\Delta_h}^2$.}
\label{F:limits}
\end{figure}

We assume that astrometric measurements of the asteroid are made at two points A and B located on the spherical Earth with the center C and radius r; see Figure~\ref{F:limits}. 
The distance $\Delta$ between A and B is so small that it is possible to neglect curvature of the Earth's surface. The observer A has a difference in distance on the geoid ($\Delta_l$) and elevation ($\Delta_h$) with respect to point B.

The difference in the astrometric positions $\Delta(\alpha,\delta)$ of the asteroid (T) with respect to the background stars (S) observed from A and B is given by a small angle $\theta$ between the directions AT and BT. Considering that the distances $x$ from A and B to the asteroid are substantially greater than the radius of the Earth, 
one finds:
\begin{equation}
\Delta=x \tan \theta
,\end{equation}
which can be simplified for $\theta$ given in arc seconds,
\begin{equation}
\Delta\approx\frac{x \;\theta} {206265}.
\end{equation}
The equality allows us to find  the upper limit of error in considering diurnal parallax with given uncertainties in the position of an observer or to set requirements to the knowledge of the position of the observer as to keep the error in reduction for parallax effect within the prescribed limits. Taking equal contributions of uncertainties in horizontal and vertical directions, one should tighten the accuracy requirement on the observer's location, i.e.,
\begin{equation}
\Delta_{l,h}\approx\frac{x\; \theta} {206265\sqrt{2}}.
\end{equation}
One can easily derive the expression for the uncertainty angle $\phi$ in angular coordinates associated with $\theta$:
\begin{equation}
\varphi\approx\frac {x \;\theta}{r},
\end{equation}
where the associated quantities are given in the same units.

\end{document}